\documentclass{article}

\begin{document}

\title{{\bf Anomalous Diffusion and Quantum Interference Effect in 
Nano-scale Periodic Lorentz Gas}}

\author{Shiro Kawabata \\
\small{Physical Science Division, Electrotechnical Laboratory,} \\ \small{1-1-4 Umezono, Tsukuba, 
Ibaraki 305-8568, Japan} \\ \small{E-mail:
shiro@etl.go.jp}}
\maketitle

\begin{abstract}
Recent advances in submicrometer technology have made it possible to confine
the two-dimensional electron gas into high-mobility semi-conductor heterostructures. Such structure with a lattice of electron-depleted circular obstacles are called
quantum antidot lattices, or quantum Lorentz gas systems.
By using the semiclassical scattering theory, we show that quantum interference in finite-size open Lorentz gas systems is expected to reflect the difference between 
normal and anomalous diffusions, i.e., L\'evy flights.\\
\end{abstract}

Recent advances in submicrometer technology have made it possible to confine
the two-dimensional electron gas into high-mobility semi-conductor heterostructures. Such
structure with a lattice of electron-depleted circular obstacles are called
$antidot$ $lattices$~\cite{rf:Weiss} and can be regarded as nano-scale periodic Lorentz gas.

In the hexagonal lattice Lorentz gas with $R/L < \sqrt{3} / 4$ 
($R$ and $L$ are the radius of the circle and the width of the unit-cell, respectively), 
there exist arbitrarily long paths along which classical 
particles can move freely without touching the hard discs (antidots). 
Thus, the diffusion in this system becomes anomalous and can be modeled 
by L\'evy flights~\cite{rf:Levy}.
In the case of sufficiently 
large radius compared to the lattice constant, $i.e.,$ $R/L > \sqrt{3} / 4$, on the other hand, 
collisionless long
trajectories can no longer exist and the diffusion becomes normal.
Therefore, we can expect that 
the quantum interference between electron paths in these systems is expected to reflect the difference between 
normal and anomalous diffusions.
In this paper, we shall investigate the {\it anomalous diffusion of quantum particles} 
in finite-size Lorentz gas
attached to the lead wires by use of the semi-classical theory.

The quantum-mechanical conductance is related to the transmission 
amplitude $t_{n,m}$ by the Landauer formula ~\cite{rf:Landauer},
\begin{equation}
g  =   \frac{e^2}{\pi \hbar} \sum_{n,m=1}^{N_M} \left| t_{n,m} \right|^2 
,
\label{eqn:a0}
\end{equation}
where $N_M$ is the number of the mode in the lead wire.
$t_{n,m}$ is exactly given by a double integral of the retarded Green's function $G$ at the Fermi energy ~\cite{rf:FL},
\begin{equation}
    t_{n,m}  =   c_{n,m} 
    \int dy \int dy' \psi_n^*(y') \psi_m(y) 
  G(y',y,E_F). 
  \label{eqn:a1}
\end{equation}
In eq.(\ref{eqn:a1}) $c_{n,m} \equiv i \hbar \sqrt{\upsilon_n \upsilon_m }$, where 
\(\upsilon_m(\upsilon_n)\) is the longitudinal velocity, 
and \(\psi_m(\psi_n)\) is 
transverse wave function for the mode $m(n)$. 
To approximate $t_{n,m}$ we replace $G$ by its semi-classical 
Feynman path-integral expression ~\cite{rf:Gutzwiller},
\begin{eqnarray}
G^{sc}(y',y,E) = \frac {2 \pi} {(2 \pi i \hbar)^{3/2}} \sum_{s(y,y')} 
  \sqrt{D_s} 
  \exp \left[ 
                            \frac i {\hbar} S_s (y',y,E) - i \frac \pi {2} \mu_s
                       \right],
  \label{eqn:a2}
\end{eqnarray}
where $S_s$ is the action integral along classical path $s$, 
 \( D_s = ( \upsilon_F \cos{\theta'})^{-1} \left| ( \partial \theta /\partial y'  )_y \right| \)
, \( \theta \) (\( \theta' \)) is the incoming (outgoing) angle, and \( \mu_s \) is the Maslov index.
Substituting eq. (\ref{eqn:a2}) into 
eq. (\ref{eqn:a1}) and using the dwelling time distribution, we finally obtain the correlation function for the $g(k)$ as
\begin{eqnarray}
    C(\Delta k)  &\equiv&   \left<  \delta g (k) \delta g (k+\Delta k) \right>_k \nonumber\\
                 &=&   \frac{e^4} {16 \pi^2 \hbar^2} \frac{1} {1+ \left( l_0 \Delta k\right)^2},
  \label{eqn:a3}
\end{eqnarray}
where $\delta g = g-g_{cl}$ ($g_{cl}$ is the classical conductance) and $l_0$ is the typical 
dwelling length in the Lorentz gas \cite{rf:Kawabata}.

From the classical simulations, we have confirmed that $l_0$ damps exponentially fast with decreasing $R/L$.
Therefore in the case that $R/L \ll(>)\sqrt{3} / 4$, $g(k)$ oscillates regularly(irregularly).
This result means that we can experimentally observe the quantum signature of anomalous diffusion 
in the Lorentz gas 
through quantum interference effects.

\end{document}